\documentclass[conference]{IEEEtran}
\IEEEoverridecommandlockouts
\usepackage{cite}
\usepackage{amsmath,amssymb,amsfonts}
\usepackage{algorithmic}
\usepackage{graphicx}
\usepackage{textcomp}
\usepackage{xcolor}
\def\BibTeX{{\rm B\kern-.05em{\sc i\kern-.025em b}\kern-.08em
    T\kern-.1667em\lower.7ex\hbox{E}\kern-.125emX}}

\DeclareMathOperator*{\argmax}{arg\,max}
\DeclareMathOperator*{\argmin}{arg\,min}
\usepackage{mathtools}

\usepackage{steinmetz}

\usepackage{blindtext}
\usepackage{caption}
\usepackage{subcaption}

\newtheorem{lemma}{Lemma}

\newtheorem{proposition}{Proposition}

\begin{document}

\title{A Low-Complexity Design for Rate-Splitting Multiple Access in Overloaded MIMO Networks}

\author{
\IEEEauthorblockN{Onur~Dizdar,
        Ata~Sattarzadeh,
        Yi~Xien~Yap,
        and~Stephen~Wang}
\IEEEauthorblockA{Research \& Technology Group, VIAVI Solutions, London, UK\\
Email: \{onur.dizdar, ata.sattarzadeh, yap.yixien, stephen.wang\}@viavisolutions.com}

%\IEEEauthorblockN{Onur Dizdar}
%\IEEEauthorblockA{\textit{Research \& Technology Group} \\
%\textit{VIAVI Solutions}\\
%London, UK \\
%onur.dizdar@viavisolutions.com}
%\and
%\IEEEauthorblockN{Ata Sattarzadeh}
%\IEEEauthorblockA{\textit{Research \& Technology Group} \\
%\textit{VIAVI Solutions}\\
%London, UK \\
%ata.sattarzadeh@viavisolutions.com}
%\and
%\vspace{0.3cm}
%\IEEEauthorblockN{Stephen Wang}
%\IEEEauthorblockA{\textit{Research \& Technology Group} \\
%\textit{VIAVI Solutions}\\
%London, UK \\
%stephen.wang@viavisolutions.com}
%\vspace{-0.7cm}
%\and
%\IEEEauthorblockN{Yi Xien Yap}
%\IEEEauthorblockA{\textit{Research \& Technology Group} \\
%\textit{VIAVI Solutions}\\
%London, UK \\
%yap.yixien@viavisolutions.com}
}

\maketitle

\begin{abstract}
Rate-Splitting Multiple Access (RSMA) is a robust multiple access scheme for multi-antenna wireless networks. In this work, we study the performance of RSMA in downlink overloaded networks, where the number of transmit antennas is smaller than the number of users. We provide analysis and closed-form solutions for optimal power and rate allocations that maximize max-min fairness when low-complexity precoding schemes are employed. The derived closed-form solutions are used to propose a low-complexity RSMA system design for precoder selection and resource allocation for arbitrary number of users and antennas under perfect Channel State Information at the Transmitter (CSIT). We compare the performance of the proposed design with benchmark designs based on Space Division Multiple Access (SDMA) to show that the proposed low-complexity RSMA design achieves a significantly higher performance gain in overloaded networks.
\end{abstract}

\begin{IEEEkeywords}
Rate-splitting, overloaded networks, multi-antenna communication, max-min fairness.
\end{IEEEkeywords}

\section{Introduction}
The demand for increased connectivity and user density has become a key performance indicator for state-of-the-art communications standards, and the increasing trend of these parameters is expected to continue for the foreseeable future \cite{rajatheva_2020}. 
%A ubiquitous coverage over a massive number of devices is necessary to realize the envisioned wireless networks of the future. 
%This brings forth several challenges in the system design, such as, user scheduling, interference management, and channel state information (CSI) acquisition from each device to be served. 
The trend of increasing user density imply that overloaded systems will be a common occurrence in the next generation networks. In overloaded systems, the number of devices to be served is larger than the number of transmit antennas. Overloaded systems have become common especially in emerging application scenarios, such as, massive Internet-of-Things (IoT) and satellite communications \cite{mao_2021, yin_2021, yin_2021_2, yin_2023, si_2022}. 
Such systems suffer from increased overhead and complexity due to signalling, user scheduling, interference management, and channel state information (CSI) acquisition, which brings a significant burden to the systems with increasing number of users. 
As the multi-user interference cannot be fully cancelled even with perfect Channel State Information at the Transmitter (CSIT), the Degree-of-Freedom (DoF) of overloaded Space Division Multiple Access (SDMA), which is the key enabling technology to support connectivity in state-of-the-art communication systems, collapse to zero, leading to a saturating achievable rate with increasing SNR \cite{clerckx_2021, yin_2021, joudeh_2017}. 

In this work, we consider Rate-Splitting Multiple Access (RSMA) to perform downlink multiple-access communications in overloaded networks. 
RSMA is a multiple access technique for multi-antenna BC that relies on linearly precoded Rate-Splitting at the transmitter and Successive Interference Cancellation (SIC) at the receivers. 
It has been shown that RSMA encapsulates and outperforms existing multiple access techniques, such as SDMA and NOMA, and can address the problems in state-of-the-art communication systems by its interference management capabilities \cite{clerckx_2021, mao_2018, clerckx_2016, mao_2019, dizdar_2020,  mao_2022_2, clerckx_2023, mishra_2022_4}. 
The max-min rate performance of RSMA in overloaded networks has been studied in \cite{yin_2021, yin_2021_2, yin_2023, si_2022, joudeh_2017, xu_2023, mao_2020, yu_2019, yalcin_2020, chen_2020, chen_2021, chen2_2020}. However, the aforementioned works formulate optimization problems for joint precoder design, power and rate allocation, which can only be solved by interior-point methods, limiting their application to practical systems.
%In order to obtain a practical design, it is essential to reduce computational complexity of the proposed design by using low-complexity precoders, such as, Zero-Forcing (ZF) and Maximum Ratio Transmission (MRT). 
%For underloaded systems, low-complexity precoders have been employed in \cite{hao_2015, dai_2017, dizdar_2021, dai_2016, papazafeiropoulos_2018, clerckx_2020} for analysis and to obtain low-complexity RSMA designs. 
%The common target in these studies is to obtain a low-complexity power allocation method to distribute power between the common and private streams that maximizes the sum-rate.
%Consequently, only a single power allocation parameter needs to be optimized once precoders are set. 

We propose a low-complexity design for RSMA in downlink overloaded MIMO networks. First, we formulate a max-min fairness problem over the ergodic rates. We adopt low-complexity ZF and MRT precoders for the private streams to limit the design space of the problem. 
The resulting formulations involve two optimization variables, namely rate and power allocation parameters.%, as opposed to the works for underloaded scenarios. 
%Depending on the type of precoders employed for the private streams, we transform the formulated problem into two problems of different forms, leading to different system designs, which involve two optimization variables, namely rate and power allocation parameters, as opposed to the works for underloaded scenarios. 
We derive expressions and bounds for the approximate ergodic rates to be used for solving the formulated problems. We obtain closed-form solutions for rate and power allocation in terms of the number of users, transmit antennas, total transmit power, and user channel disparity. Finally, we propose a low-complexity RSMA system design for overloaded networks based on the derived solutions. 
To the best of our knowledge, this is the first paper analysing RSMA in overloaded networks with low-complexity precoders and proposing a low-complexity solution for precoder selection, rate and power allocation for max-min fairness.

%\textit{Notation:} Vectors are denoted by bold lowercase letters. 
%The operations $|.|$ and $||.||$ denote the cardinality of a set or absolute value of a scalar and $l_{2}$-norm of a vector, respectively. 
%The operation $\mathbf{a}^{H}$ denotes the Hermitian transpose of a vector $\mathbf{a}$.  
%$\mathcal{CN}(0,\sigma^{2})$ denotes the Circularly Symmetric Complex Gaussian distribution with zero mean and variance $\sigma^{2}$. 
%$\mathbf{I}$ denotes the identity matrix. $\left \lfloor{.}\right \rceil $ denotes the round operation. 
%Logarithms are natural logarithms, {\sl i.e.,} $\log(.)=\log_{e}(.)$, unless stated otherwise. $\mathrm{Gamma}(D,\theta)$ represents the Gamma distribution with shape $D$ and scale $\theta$ and $\chi^{2}_{k}$ represents the chi-squared distribution with degrees of freedom $k$. For any complex $x$ with a positive real part, $\Gamma(x)$ is the gamma function and $\Gamma^{\prime}(x)$ is the derivative of $\Gamma(x)$ with respect to $x$ and $\psi(x)=\frac{\Gamma^{\prime}(x)}{\Gamma(x)}$. 

\section{System Model}
\label{sec:system}
\begin{figure}[t!]
	\centerline{\includegraphics[width=3.0in,height=3.0in,keepaspectratio]{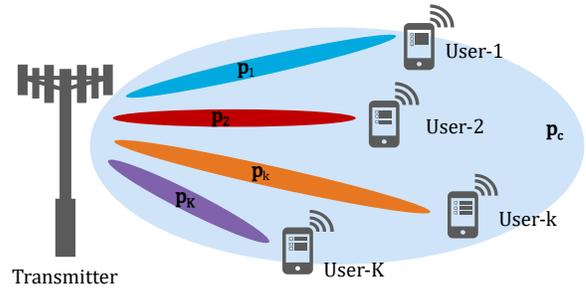}}
	\caption{Transmission model of $K$-user 1-layer RSMA.}
	\label{fig:system}
	\vspace{-0.6cm}
\end{figure}
We consider the system model in Fig.~\ref{fig:system} consisting of one transmitter with $N$ transmit antennas and $K$ single-antenna users indexed by $\mathcal{K}=\{1,\ldots,K\}$, with $N < K$. We employ 1-layer RSMA for downlink multi-user transmission, which performs message splitting at the transmitter to obtain a single common stream for all users. The message for user-$k$, denoted by $W_{k}$, is split into two independent parts, namely common part $W_{c,k}$ and private part $W_{p,k}$, $\forall k\in\mathcal{K}$. 
The common parts of all user messages are combined into a single common message, denoted by $W_{c}$. The messages $W_{c}$ and $W_{p,k}$ are independently encoded into streams $s_{c}$ and $s_{k}$, respectively. After linear precoding, signal to be transmitted is written as
\vspace{-0.1cm}
\begin{align}	
\mathbf{x}=\sqrt{P(1-t)}\mathbf{p}_{c}s_{c}+\sqrt{Pt}\sum_{k \in \mathcal{K}}\sqrt{\mu_{k}}\mathbf{p}_{k}s_{k},
	\label{eqn:1}
\end{align} 
where $||\mathbf{p}_{c}||^{2}=1$, $||\mathbf{p}_{k}||^{2}=1$. Power allocation coefficient $0 \leq t \leq 1$ determines the distribution of power among the common and the private precoders and $0 \leq \mu_{k} \leq 1$ determines the portion of the private streams' power allocated to user-$k$. 
Setting $t=1$ turns off the common stream and RSMA boils down to SDMA.
The signal received by user-$k$ is written as
\vspace{-0.1cm}
\begin{align}
	y_{k}=\sqrt{v_{k}}\mathbf{h}_{k}^{H}\mathbf{x}+n_{k}, \quad k\in\mathcal{K},  
\end{align}
where $\mathbf{h}_{k} \in \mathbb{C}^{N}$ is the channel vector representing small scale fading with i.i.d. entries distributed by $\mathcal{CN}(0,1)$, $0 < v_{k} \leq 1$ is the channel coefficient representing large scale fading between the transmitter and user-$k$, and $n_{k} \sim \mathcal{CN}(0,1)$ is the Additive White Gaussian Noise (AWGN) at user-$k$. 

Each user receives its signal by first decoding the common stream to obtain $\widehat{W}_{c}$ and extracting $\widehat{W}_{c,k}$ from $\widehat{W}_{c}$. Then, SIC is performed by reconstructing the common stream using $\widehat{W}_{c}$ and removing the reconstructed signal from the received signal. Finally, each user decodes its private stream and obtains $\widehat{W}_{p,k}$ by treating the interference from other users' private streams as noise.
The overall decoded message of user-$k$, $\widehat{W}_{k}$, is reconstructed by combining $\widehat{W}_{c,k}$ and $\widehat{W}_{p,k}$.
The Signal-to-Interference-plus-Noise Ratio (SINR) and the ergodic rates for the common and private streams at user-$k$ are written as\vspace{-0.1cm}
\begin{align}
	&\gamma_{c,k}\hspace{-0.1cm}=\hspace{-0.1cm}\frac{P(1-t)v_{k}|\mathbf{h}_{k}^{H}\mathbf{p}_{c}|^{2}}{1\hspace{-0.1cm}+\hspace{-0.1cm}Ptv_{k}\hspace{-0.1cm}\sum_{j \in \mathcal{K}}\mu_{j}\hspace{-0.03cm}|\mathbf{h}_{k}^{H}\hspace{-0.07cm}\mathbf{p}_{j}|^{2}}, 
	\gamma_{k}\hspace{-0.1cm}=\hspace{-0.1cm}\frac{\mu_{k}Ptv_{k}|\mathbf{h}_{k}^{H}\mathbf{p}_{k}|^{2}}{1\hspace{-0.1cm}+\hspace{-0.1cm}Ptv_{k}\hspace{-0.1cm}\sum_{\substack{j \in \mathcal{K} \\ j \neq k}}\mu_{j}\hspace{-0.03cm}|\mathbf{h}_{k}^{H}\hspace{-0.07cm}\mathbf{p}_{j}|^{2}},\nonumber\\
&R_{c}(t)\hspace{-0.1cm}=\hspace{-0.1cm}\mathbb{E}\left\lbrace \hspace{-0.1cm}\log_{2}\hspace{-0.1cm}\left(\hspace{-0.1cm} 1\hspace{-0.1cm}+\hspace{-0.1cm}\min_{k\in\mathcal{K}}\gamma_{c,k} \right) \hspace{-0.15cm}\right\rbrace\hspace{-0.1cm}, \ 
	R_{k}(t)\hspace{-0.1cm}=\hspace{-0.1cm}\mathbb{E}\left\lbrace \log_{2}\hspace{-0.05cm}\left( 1\hspace{-0.1cm}+\hspace{-0.1cm} \gamma_{k} \right)\right\rbrace\hspace{-0.1cm},
	\label{eqn:rates}
\end{align}
where the expectations are defined over the user channels $\mathbf{h}_{k}$.
The precoders $\mathbf{p}_{c}$ and $\mathbf{p}_{k}$ are calculated based on perfect CSIT. 

\section{Problem Formulation}
\label{sec:formulation}
We write a max-min fairness optimisation problem as \vspace{-0.1cm}%to ensure a non-zero ergodic rate for each user as\vspace{-0.2cm}
\begin{align}
	\max_{\mathbf{P}, \mathbf{c}, \boldsymbol{\mu}, t} \min_{k \in \mathcal{K}} C_{k}(t)+R_{k}(t), 
	\label{eqn:problem}
\end{align}
where $C_{k}(t)$ is the portion of the common stream rate allocated to user-$k$, $\mathbf{c}=[C_{1}, C_{2}, \ldots, C_{K}]$, $\mathbf{P}=[\mathbf{p}_{c}, \mathbf{p}_{1}, \mathbf{p}_{2}, \ldots, \mathbf{p}_{K}]$, and $\boldsymbol{\mu}=[\mu_{1}, \mu_{2}, \ldots, \mu_{K}]$. 
Our target is to obtain a low-complexity design as a solution of problem \eqref{eqn:problem}. We start simplifying the solution by limiting the design space for optimal precoders. Accordingly, we consider two types of precoders, specifically ZF and MRT, to be used for transmission of private streams. The type of employed precoders affect the expressions for the rates $C_{k}(t)$ and $R_{k}(t)$, which in turn affects the system design and resource allocation. 
%Limiting the design space for precoders allows us to formulate separate problem formulations for different types of precoders employed. 
Let us define $C^{\mathrm{ZF}}_{k}(t)$ and $R^{\mathrm{ZF}}_{k}(t)$ to denote the portion of the common stream rate and the private stream rate for user-$k$, respectively, when ZF precoders are used for private streams, and $C^{\mathrm{MRT}}_{k}(t)$ and $R^{\mathrm{MRT}}_{k}(t)$ when MRT precoders are used. Then, we reformulate \eqref{eqn:problem} as \vspace{-0.1cm}
\begin{align}
	\max \hspace{-0.1cm}\bigg(\hspace{-0.15cm}\max_{\mathbf{c}, \boldsymbol{\mu}, t} \min_{k \in \mathcal{K}} C^{\mathrm{ZF}}_{k}\hspace{-0.08cm}(t)\hspace{-0.1cm}+\hspace{-0.1cm}R^{\mathrm{ZF}}_{k}\hspace{-0.08cm}(t), \max_{\mathbf{c}, \boldsymbol{\mu}, t} \min_{k \in \mathcal{K}} C^{\mathrm{MRT}}_{k}\hspace{-0.08cm}(t)\hspace{-0.1cm}+\hspace{-0.1cm}R^{\mathrm{MRT}}_{k}\hspace{-0.08cm}(t)\hspace{-0.1cm}\bigg).
	\label{eqn:problem2}
\end{align}

\subsection{ZF Precoding}
\label{sec:formulation_ZF}
The DoF analysis in \cite{clerckx_2021} reveals that the optimal max-min DoF is achieved by serving $N$ users by private streams with ZF precoding and a portion of the common stream, while the remaining $K-N$ users are served only by a portion of the common stream. To this end, let us define the user groups $\mathcal{G}_{1}$ and $\mathcal{G}_{2}$, satisfying $|\mathcal{G}_{1}|=N$, $|\mathcal{G}_{2}|=K-N$, $\mathcal{G}_{1}\cap\mathcal{G}_{2}=\emptyset$, and, $\mathcal{G}_{1}\cup\mathcal{G}_{2}=\mathcal{K}$. $\mathcal{G}_{1}$ denotes the index set for users which are served by private and common streams, while $\mathcal{G}_{2}$ is the index set for users served by common stream only. 
We assume equal power allocation among the private streams, such that, $\mu_{k}=\frac{1}{N}$, if $k \in \mathcal{G}_{1}$ and $\mu_{k}=0$ if $k \in \mathcal{G}_{2}$.
%We also assume that the rate of the common stream is distributed equally among the users in $\mathcal{G}_{1}$ and $\mathcal{G}_{2}$. 
Let us define the rate allocation coefficient, $0\leq \beta < \frac{1}{N}$, to denote the proportion of the common stream rate allocated to each user in $\mathcal{G}_{1}$.
We reformulate the first term in \eqref{eqn:problem2} as \vspace{-0.1cm}
\begin{align}
	\max_{\substack{\beta \in [0,1/N),\\ t\in [0,1)}}\min_{k \in \mathcal{G}_{1}} \left\{\beta R^{\mathrm{ZF}}_{c}(t)+R^{\mathrm{ZF}}_{k}(t), \frac{1-N\beta}{K-N}R^{\mathrm{ZF}}_{c}(t)\right\}.
	\label{eqn:ZFproblem_1}
\end{align}
The first term in parenthesis in \eqref{eqn:ZFproblem_1} represents the total rate allocated to users in $\mathcal{G}_{1}$ and the second term represents the rate allocated to users in $\mathcal{G}_{2}$. Defining $\tilde{k}=\argmin_{k \in \mathcal{G}_{1}}R^{\mathrm{ZF}}_{\tilde{k},p}(t)$, \eqref{eqn:ZFproblem_1} can be expressed in a simpler form as\vspace{-0.1cm}
\begin{align}
	\max_{\substack{\beta \in [0,1/N),\\ t\in [0,1)}}\min \left(\beta R^{\mathrm{ZF}}_{c}(t)+R^{\mathrm{ZF}}_{\tilde{k}}(t), \frac{1-N\beta}{K-N}R^{\mathrm{ZF}}_{c}(t)\right).
	\label{eqn:ZFproblem_2}
\end{align}

Let us denote the ZF precoder for user-$k$ as $\mathbf{p}^{\mathrm{ZF}}_{k}$, $\forall k \in \mathcal{G}_{1}$, such that, $\mathbf{h}^{H}_{j}\mathbf{p}^{\mathrm{ZF}}_{k}=0$ for $\forall j,k \in \mathcal{G}_{1}$ and $j \neq k$. Accordingly, the rate expressions in \eqref{eqn:ZFproblem_2} can be written explicitly as \vspace{-0.1cm}
\begin{align}
	R^{\mathrm{ZF}}_{c}(t)&=\mathbb{E}\left\lbrace \hspace{-0.1cm}\log_{2}\hspace{-0.1cm}\left(\hspace{-0.1cm} 1\hspace{-0.1cm}+\hspace{-0.1cm}\min_{k\in\mathcal{K}}\frac{P(1-t)v_{k}|\mathbf{h}_{k}^{H}\mathbf{p}_{c}|^{2}}{1+\frac{Pt}{N}v_{k}\sum_{j \in \mathcal{G}_{1}}|\mathbf{h}_{k}^{H}\mathbf{p}^{\mathrm{ZF}}_{j}|^{2}} \right) \hspace{-0.1cm}\right\rbrace, \nonumber \\
	R^{\mathrm{ZF}}_{\tilde{k}}(t)\hspace{-0.1cm}&=\hspace{-0.1cm}\mathbb{E}\left\lbrace \log_{2}\hspace{-0.05cm}\left( 1\hspace{-0.1cm}+\hspace{-0.1cm} \frac{Pt}{N}v_{\tilde{k}}|\mathbf{h}_{\tilde{k}}^{H}\mathbf{p}^{\mathrm{ZF}}_{\tilde{k}}|^{2} \right) \right\rbrace.
	\label{eqn:rates_ZF}
\end{align}

\subsection{MRT Precoding}
\label{sec:formulation_MRT}
Next, we propose a design where each user is served by a private stream encoded by MRT precoders and a portion of the common stream. Therefore, users are not divided into two groups. We assume equal power allocation among the private streams, such that, $\mu_{k}=\frac{1}{N}$, $\forall k \in \mathcal{K}$ and that the common rate is distributed equally among all users. The rate expression for each user is identical, and the second term in \eqref{eqn:problem2} becomes\vspace{-0.1cm}
\begin{align}
	\max_{t\in (0,1]} \ \frac{1}{K}R^{\mathrm{MRT}}_{c}(t)+R^{\mathrm{MRT}}_{\hat{k}}(t).
	\label{eqn:MRTproblem}
\end{align}
where $\hat{k}=\argmin_{k \in \mathcal{K}}R^{\mathrm{MRT}}_{k}(t)$. 
Defining the MRT precoder for user-$k$ as $\mathbf{p}^{\mathrm{MRT}}_{k}=\frac{\widehat{\mathbf{h}}_{k}}{|\widehat{\mathbf{h}}_{k}|}$, the rates in \eqref{eqn:MRTproblem} are written as \vspace{-0.1cm}
\begin{align}
	R^{\mathrm{MRT}}_{c}(t)&\hspace{-0.1cm}=\hspace{-0.1cm}\mathbb{E}\hspace{-0.1cm}\left\lbrace \hspace{-0.1cm}\log_{2}\hspace{-0.1cm}\left(\hspace{-0.1cm} 1\hspace{-0.1cm}+\hspace{-0.1cm}\min_{k\in\mathcal{K}}\frac{P(1-t)v_{k}|\mathbf{h}_{k}^{H}\mathbf{p}_{c}|^{2}}{1+\frac{Pt}{K}v_{k}\sum_{j \in \mathcal{K}}|\mathbf{h}_{k}^{H}\mathbf{p}^{\mathrm{MRT}}_{j}|^{2}} \right) \hspace{-0.1cm}\right\rbrace, \nonumber \\
	R^{\mathrm{MRT}}_{\tilde{k}}(t)\hspace{-0.1cm}&=\hspace{-0.1cm}\mathbb{E}\hspace{-0.1cm}\left\lbrace \hspace{-0.1cm}\log_{2}\hspace{-0.1cm}\left(\hspace{-0.1cm}1\hspace{-0.1cm}+\hspace{-0.1cm} \frac{\frac{Pt}{K}v_{\tilde{k}}|\mathbf{h}_{\tilde{k}}^{H}\mathbf{p}^{\mathrm{MRT}}_{\tilde{k}}|^{2}}{1+\frac{Pt}{K}v_{\tilde{k}}\sum_{\substack{j \in \mathcal{K}, \\ j \neq \tilde{k}}}|\mathbf{h}_{\tilde{k}}^{H}\mathbf{p}^{\mathrm{MRT}}_{j}|^{2}} \hspace{-0.1cm}\right) \hspace{-0.2cm}\right\rbrace\hspace{-0.1cm}.\hspace{-0.1cm}
	\label{eqn:rates_MRT}
\end{align}
With the simplified formulations in \eqref{eqn:ZFproblem_2} and \eqref{eqn:MRTproblem}, \eqref{eqn:problem} becomes\vspace{-0.1cm}
\begin{align}
	\max&\bigg(\hspace{-0.1cm}\max_{\substack{\beta \in [0,1/N),\\ t\in [0,1)}}\hspace{-0.2cm}\min \left(\hspace{-0.1cm}\beta R^{\mathrm{ZF}}_{c}(t)\hspace{-0.1cm}+\hspace{-0.1cm}R^{\mathrm{ZF}}_{\tilde{k}}(t), \frac{1\hspace{-0.1cm}-\hspace{-0.1cm}N\beta}{K\hspace{-0.1cm}-\hspace{-0.1cm}N}R^{\mathrm{ZF}}_{c}(t)\hspace{-0.1cm}\right)\hspace{-0.1cm}, \nonumber \\
	&\max_{t\in (0,1]} \frac{1}{K}R^{\mathrm{MRT}}_{c}(t)\hspace{-0.1cm}+\hspace{-0.1cm}R^{\mathrm{MRT}}_{\hat{k}}(t)\hspace{-0.1cm}\bigg).
	\label{eqn:problem3}
\end{align}

\section{Expressions and Bounds for Achievable Rates}
\label{sec:boundRate}
In this section, we provide lower bounds for the ergodic rates in \eqref{eqn:rates_ZF} and \eqref{eqn:rates_MRT} as the next step towards finding closed-form solutions for optimal $t$ and $\beta$. 
We start by providing several useful properties. 
$\mathbf{p}^{\mathrm{ZF}}_{k}$ is orthogonal to $\mathbf{h}_{j}$, $\forall k,j \in \mathcal{G}_{1}$, $k \neq j$, which reduces the degrees of freedom at the transmitter to $N-|\mathcal{G}_{1}|+1=1$. Accordingly, 
$|\sqrt{2}\mathbf{h}_{k}^{H}\mathbf{p}^{\mathrm{ZF}}_{k}|^{2}\sim \chi^{2}_{2}$ or $|\mathbf{h}_{k}^{H}\mathbf{p}^{\mathrm{ZF}}_{k}|^{2}\sim \mathrm{Gamma}(1,1)$\cite{jindal_2011}. %\cite{jindal_2011, jaramilloramirez_2015}. 
For MRT precoding, we assume that user channels are i.i.d., so that, $|\mathbf{h}_{k}^{H}\mathbf{p}^{\mathrm{MRT}}_{k}|^{2}\sim \mathrm{Gamma}(N,1)$ and $|\mathbf{h}_{j}^{H}\mathbf{p}^{\mathrm{MRT}}_{k}|^{2}\sim \mathrm{Gamma}(1,1)$.

We note that we skip some details in the following derivations for the sake of brevity. %Please refer to \cite{xxx} for a more comprehensive study on the derivations.

\subsection{Rates with ZF Precoding}
\label{sec:ratesZFprecoding}
The following propositions give lower bounds for the approximations of the ergodic rates with ZF precoding. 
\begin{proposition}
Let $\widetilde{R}^{\mathrm{ZF}}_{k}(t)$ be an approximation to $R^{\mathrm{ZF}}_{k}(t)$. Then, the following relation holds.\vspace{-0.1cm}
\begin{align}
	\widetilde{R}^{\mathrm{ZF}}_{k}(t) \geq \log_{2}\hspace{-0.1cm}\left(\hspace{-0.1cm}1+\frac{Pt}{N}v_{k}e^{\psi(1)}\hspace{-0.1cm}\right).
	\label{eqn:prop2}
\end{align}
\end{proposition}
\quad\textit{Proof:} The proof follows from \cite[Proposition 1]{dizdar_2021}. \hspace{1.0cm}  $\blacksquare$
%The bound \eqref{eqn:prop1} contains the generalized exponential-integral function, which makes it challenging for derivations. 
%We provide a simplified expression for \eqref{eqn:prop1} under the assumption $Pt \rightarrow \infty$ to be used in derivations.

\begin{proposition}
Let $\widetilde{R}^{\mathrm{ZF}}_{c}(t)$ be an approximation to $R^{\mathrm{ZF}}_{c}(t)$. The following relation holds for $Pt \rightarrow \infty$\vspace{-0.1cm}
\begin{align}
	\widetilde{R}^{\mathrm{ZF}}_{c}(t)\geq\log_{2}\left(1-\rho^{\mathrm{ZF}}\hspace{-0.1cm}+\hspace{-0.1cm}\frac{\rho^{\mathrm{ZF}}}{t}\right)\hspace{-0.1cm},
	\label{eqn:lemma1}
\end{align}
where $\rho^{\mathrm{ZF}}\triangleq\frac{N}{\lfloor N(K-N+1)\rceil-1}e^{-\gamma-\frac{1}{2\left(\lfloor N(K-N+1)\rceil-1\right)}}$ and $\gamma\approx0.577$ is the Euler-Mascheroni constant.
\end{proposition}
\quad\textit{Proof:} See Appendix~\ref{appendix:prop1}. \hspace{4.6cm}$\blacksquare$

\subsection{Rates with MRT Precoding}
\label{sec:ratesMRTprecoding}
%We define the r.v.s $X^{\mathrm{MRT}}_{k}=\frac{v_{k}|\mathbf{h}_{k}^{H}\mathbf{p}^{\mathrm{MRT}}_{k}|^{2}}{1+\frac{Pt}{K}v_{k}\sum_{\substack{j \in \mathcal{K} \\ j \neq k}}|\mathbf{h}_{k}^{H}\mathbf{p}^{\mathrm{MRT}}_{j}|^{2}}$, $Y^{\mathrm{MRT}}\hspace{-0.15cm}=\hspace{-0.1cm}\min_{k\in\mathcal{K}}Y^{\mathrm{MRT}}_{k}\hspace{-0.1cm}$, and $Y^{\mathrm{MRT}}_{k}=\hspace{-0.1cm}
%	\frac{v_{k}|\mathbf{h}_{k}^{H}\mathbf{p}_{c}|^{2}}{1+\frac{Pt}{K}v_{k}\sum_{j \in \mathcal{K}}|\mathbf{h}_{k}^{H}\mathbf{p}^{\mathrm{MRT}}_{j}|^{2}}$, such that,
%$R^{\mathrm{MRT}}_{c}(t)\hspace{-0.1cm}=\hspace{-0.1cm}\mathbb{E}\left\lbrace \log_{2}\hspace{-0.1cm}\left(1\hspace{-0.1cm}+\hspace{-0.1cm}P(1\hspace{-0.1cm}-\hspace{-0.1cm}t)Y^{\mathrm{MRT}} \right) \hspace{-0.1cm}\right\rbrace$ and $R^{\mathrm{MRT}}_{k}(t)\hspace{-0.1cm}=\hspace{-0.1cm}\mathbb{E}\left\lbrace \log_{2}\hspace{-0.05cm}\left( 1\hspace{-0.1cm}+\hspace{-0.1cm} \frac{Pt}{N}X^{\mathrm{MRT}}_{k} \right) \right\rbrace$.
The following propositions give lower bounds for the approximations of the ergodic rates with MRT precoding.
\begin{proposition}
Let $\widetilde{R}^{\mathrm{MRT}}_{k}(t)$ be an approximation to $R^{\mathrm{MRT}}_{k}(t)$. Then, the following relation holds.\vspace{-0.1cm}
\begin{align}
	&\widetilde{R}^{\mathrm{MRT}}_{k}(t) \nonumber\\
	&\geq \hspace{-0.1cm}\log_{2}\hspace{-0.1cm}\left(\hspace{-0.1cm}1\hspace{-0.1cm}+\hspace{-0.1cm}\frac{Pt}{K}v_{k}e^{\psi(N+K-1)}\hspace{-0.1cm}\right)
	\hspace{-0.1cm}-\hspace{-0.1cm}\log_{2}\hspace{-0.1cm}\left(\hspace{-0.1cm}1\hspace{-0.1cm}+\hspace{-0.1cm}\frac{Pt}{K}v_{k}(K-1)\hspace{-0.1cm}\right)\hspace{-0.1cm}.
	\label{eqn:prop4} 
\end{align}
\end{proposition}
\quad\textit{Proof:} The proof follows from \cite[Proposition 1]{dizdar_2021} by taking into account the abovementioned properties of MRT precoders, and is omitted here for the sake of brevity. \hspace{2.1cm}  $\blacksquare$

\begin{proposition}
Let $\widetilde{R}^{\mathrm{MRT}}_{c}(t)$ be an approximation to $R^{\mathrm{MRT}}_{c}(t)$. The following relation holds for $Pt \rightarrow \infty$
\begin{align}
	\widetilde{R}^{\mathrm{MRT}}_{c}(t)\geq\log_{2}\left(1-\rho^{\mathrm{MRT}}+\frac{\rho^{\mathrm{MRT}}}{t}\right)\hspace{-0.1cm},
	\label{eqn:lemma3}
\end{align}
where $\rho^{\mathrm{MRT}}\triangleq\frac{K}{\left( (N+K-1)K-1\right)}e^{-\gamma-\frac{1}{2\left((N+K-1)K-1\right)}}$. 
\end{proposition}
\quad\textit{Proof:} The proof follows from \cite[Proposition 2]{dizdar_2021} and \cite[Lemma 4]{dizdar_2021} by taking into account the abovementioned properties of MRT precoders, and is omitted here for the sake of brevity.% and refer the interested reader to \cite{XXX}.      \hspace{1.1cm}  $\blacksquare$ 

\section{Proposed Solution}
\label{sec:solution_perfectcsit}
We derive closed-form solutions for optimal $\beta$ and $t$ using the lower bounds derived in Section~\ref{sec:boundRate}. Then, we propose a low-complexity precoder and resource allocation for RSMA based on the derived closed-form solutions.

\subsection{Solution Using ZF Precoding}
\label{sec:solution_perfectcsit_ZF}
\vspace{-0.1cm}
Proposition 5 provides a condition which should be satisfied by the optimal $\beta$ and $t$, denoted by $\beta^{*}$ and $t^{*}$, respectively. 

\begin{proposition}
The maximum of \eqref{eqn:ZFproblem_2} is attained by unique $\beta^{*}$ and $t^{*}$ satisfying \vspace{-0.1cm}
\begin{align}
	\frac{1-N\beta^{*}}{K-N}R^{\mathrm{ZF}}_{c}(t^{*})=\beta^{*}R^{\mathrm{ZF}}_{c}(t^{*})+R^{\mathrm{ZF}}_{\tilde{k}}(t^{*}). 
	\label{eqn:propx}
\end{align}
\label{eqn:propp}
\end{proposition}
\vspace{-0.2cm}
\quad\textit{Proof:} See Appendix~\ref{appendix:propx}. \hspace{4.6cm}$\blacksquare$

%\begin{corollary}
%	The solution $\beta^{*}$ satisfies $0\leq\beta^{*}\leq\frac{1}{K}$.
%\end{corollary}
%\quad\textit{Proof:} The proof is straightforward once \eqref{eqn:propx} is rearranged as $\frac{1-K\beta^{*}}{K-N}R^{\mathrm{ZF}}_{c,p}(t^{*})=R^{\mathrm{ZF}}_{\tilde{k},p}(t^{*})\hspace{-0.1cm}\geq 0$. \hspace{3.4cm}$\blacksquare$

We use Proposition~5 to find $\beta^{*}$ and $t^{*}$. 
Using \eqref{eqn:prop2}, we can redefine $\tilde{k}=\argmin_{k \in \mathcal{G}_{1}}\hspace{-0.1cm}v_{k}$, such that, $R^{\mathrm{ZF}}_{\tilde{k}}(t)=\log_{2}(1+\sigma^{\mathrm{ZF}}_{\tilde{k}}t)$, where $\sigma^{\mathrm{ZF}}_{\tilde{k}}\hspace{-0.1cm}=\hspace{-0.1cm}\frac{v_{\tilde{k}}P}{N}e^{\psi(1)}$ for given $v_{k}$.
Combined with \eqref{eqn:lemma1}, we rewrite \eqref{eqn:propx} as\vspace{-0.1cm}
\begin{align}
	\frac{1-K\beta^{*}}{K-N}\log_{2}\left(\hspace{-0.03cm}1\hspace{-0.1cm}-\hspace{-0.1cm}\rho^{\mathrm{ZF}}\hspace{-0.1cm}+\hspace{-0.1cm}\rho^{\mathrm{ZF}}/t^{*}\right)=\log_{2}\left(1+\sigma^{\mathrm{ZF}}_{\tilde{k}}t^{*}\right).
	\label{eqn:perfectcsit_highsnr}
\end{align}

First, let us consider the scenario where $R^{\mathrm{ZF}}_{\tilde{k}}(t)$ is large enough, such that, $\log_{2}(1+\sigma^{\mathrm{ZF}}_{\tilde{k}}t^{*})\approx\log_{2}(\sigma^{\mathrm{ZF}}_{\tilde{k}}t^{*})$. We note $\sigma^{\mathrm{ZF}}_{\tilde{k}}>1$ in this scenario. Accordingly, the left-hand term in \eqref{eqn:perfectcsit_highsnr} is also expected to be large. Using this assumption and the fact that $1-\rho^{\mathrm{ZF}}\leq 1$, one can approximate $\log_{2}\left(1-\rho^{\mathrm{ZF}}+\rho^{\mathrm{ZF}}/t^{*}\right)\hspace{-0.1cm}\approx\hspace{-0.1cm}\log_{2}\left(\rho^{\mathrm{ZF}}/t^{*}\right)$. The resulting expression can be arranged to obtain 
\begin{align}
t^{*}=\left(\frac{(\rho^{\mathrm{ZF}})^{1-K\beta^{*}}}{(\sigma^{\mathrm{ZF}}_{\tilde{k}})^{K-N}}\right)^{\frac{1}{1-K\beta^{*}+K-N}}\triangleq t^{(1)}(\beta^{*}). 
\end{align}
Then, we can find $\beta^{*}$ by maximizing either $\frac{1-N\beta}{K-N}R^{\mathrm{ZF}}_{c}(t^{(1)}(\beta))$ or $\beta R^{\mathrm{ZF}}_{c}(t^{(1)}(\beta))+R^{\mathrm{ZF}}_{\tilde{k}}(t^{(1)}(\beta))$, since they are equal at $\beta^{*}$ from Proposition 5. We provide a useful result in Lemma 1 to find $\beta^{*}$.% by maximizing $\frac{1-N\beta}{K-N}R^{\mathrm{ZF}}_{c}(t^{(1)}(\beta))$ with respect to $\beta$.

\begin{lemma}
$\frac{1-N\beta}{K-N}\log_{2}\left(1-\rho^{\mathrm{ZF}}+\frac{\rho^{\mathrm{ZF}}}{t^{(1)}(\beta)}\right)$ is a monotonic decreasing function of $\beta$ for $0\leq \beta\leq \frac{1}{K}$ and $\sigma^{\mathrm{ZF}}_{\tilde{k}}>1$. 
\end{lemma}
\quad\textit{Proof:} See Appendix~\ref{appendix:lemmax}. \hspace{4.6cm}$\blacksquare$

Lemma 1 shows that the maximum of $\frac{1-N\beta}{K-N}R^{\mathrm{ZF}}_{c}(t^{(1)}(\beta))$ is attained by choosing $\beta$ as small as possible. Therefore, we set $\beta^{*}=0$, which means that the common stream needs to serve only the users in $G_{2}$ for optimal performance. 
%\begin{remark}
%The obtained result $\beta^{*}=0$ is coherent with the result of DoF analysis in \cite{joudeh_2017}.
%\end{remark}
We define the optimal rates and parameters for this scenario in \eqref{eqn:1_perfect}.
%One can easily see that $ t^{(1)}\geq 0$ since $\sigma^{\mathrm{ZF}}_{\tilde{k}}\geq 0$ and $\rho^{\mathrm{ZF}}\geq 0$.
\begin{table*}
\begin{align}
		&t^{(1)}\triangleq \min\left\lbrace\left(\frac{\rho^{\mathrm{ZF}}}{(\sigma^{\mathrm{ZF}}_{\tilde{k}})^{K-N}}\right)^{\frac{1}{1+K-N}}\hspace{-0.3cm},1\right\rbrace\hspace{-0.1cm}, \quad \beta^{(1)}=0, \quad r^{(1)}_{mm}\hspace{-0.1cm}\triangleq\hspace{-0.1cm}\min\hspace{-0.1cm}\left\lbrace\hspace{-0.1cm}\frac{\log_{2}\hspace{-0.1cm}\left(\hspace{-0.1cm}1-\rho^{\mathrm{ZF}}+\frac{\rho^{\mathrm{ZF}}}{t^{(1)}}\right)}{K-N}, \log_{2}\left(1\hspace{-0.05cm}+\hspace{-0.05cm}\sigma^{\mathrm{ZF}}_{\tilde{k}}t^{(1)}\right)\hspace{-0.1cm}\right\rbrace\hspace{-0.1cm}.		
		\label{eqn:1_perfect}
\end{align}
\vspace{-0.15cm}
\hrule
\vspace{-0.3cm}
\end{table*}
\begin{figure}[t!]
	\begin{subfigure}{.24\textwidth}
		\centerline{\includegraphics[width=1.9in,height=1.9in,keepaspectratio]{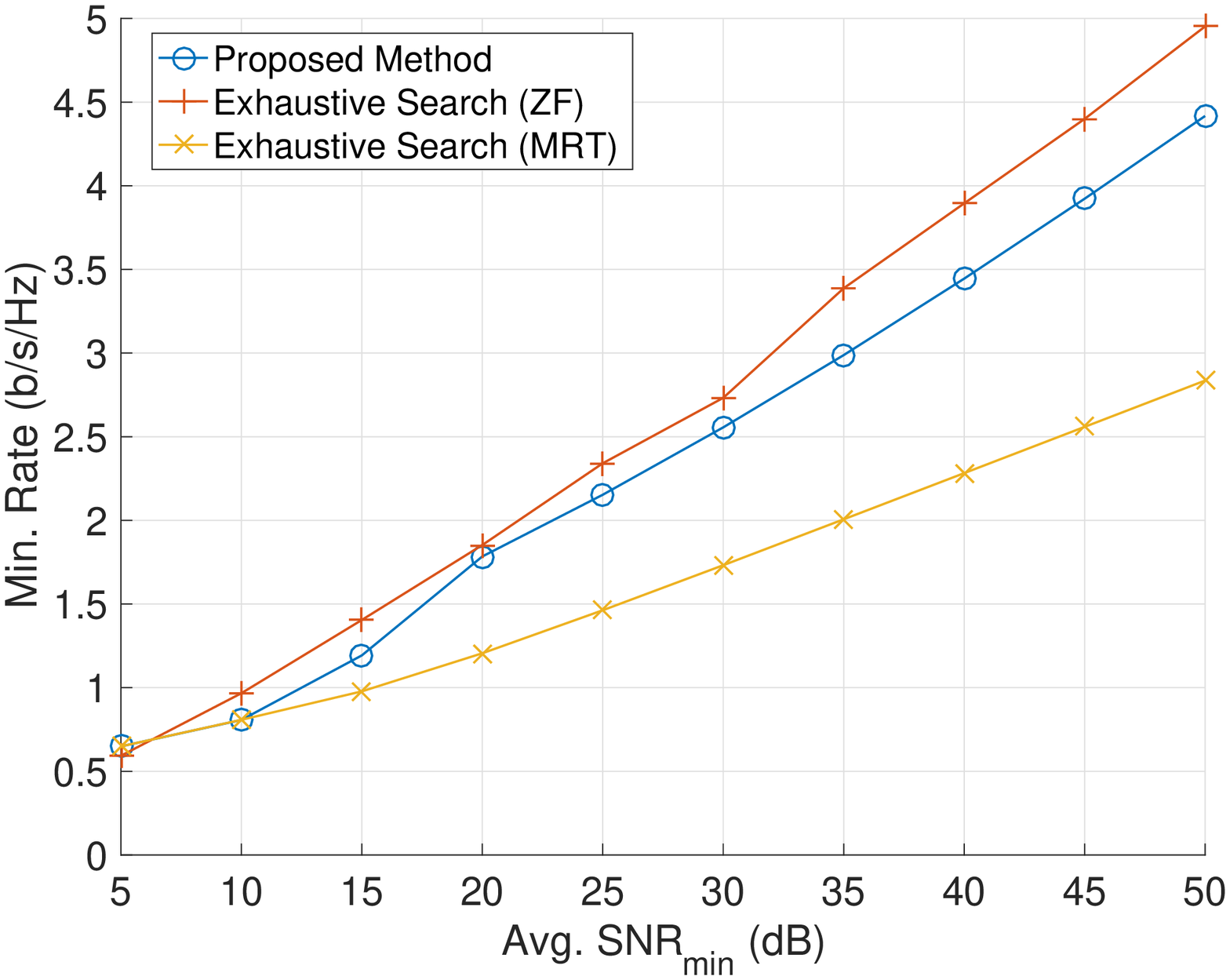}}
		\vspace{-0.15cm}
		\caption{$N=4$, $K=6$.}
		\label{fig:N4K6_PerfectCSIT_Exhaustive}
	\end{subfigure}
	\begin{subfigure}{.24\textwidth}
		\centerline{\includegraphics[width=1.9in,height=1.9in,keepaspectratio]{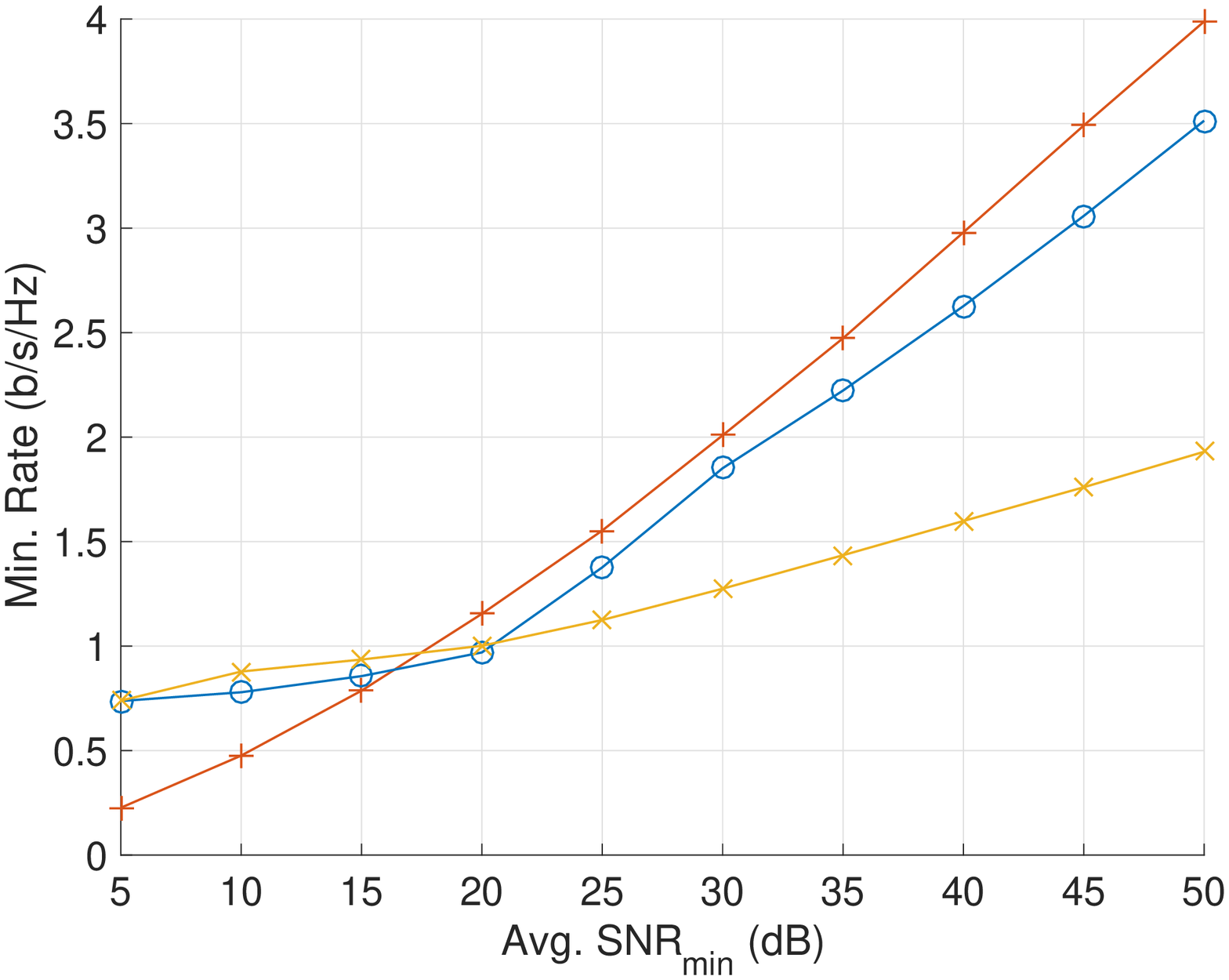}}
		\vspace{-0.15cm}
		\caption{$N=8$, $K=10$.}
		\label{fig:N16K20_PerfectCSIT_Exhaustive}
	\end{subfigure}
	\newline
	\begin{subfigure}{.24\textwidth}
		\centerline{\includegraphics[width=1.9in,height=1.9in,keepaspectratio]{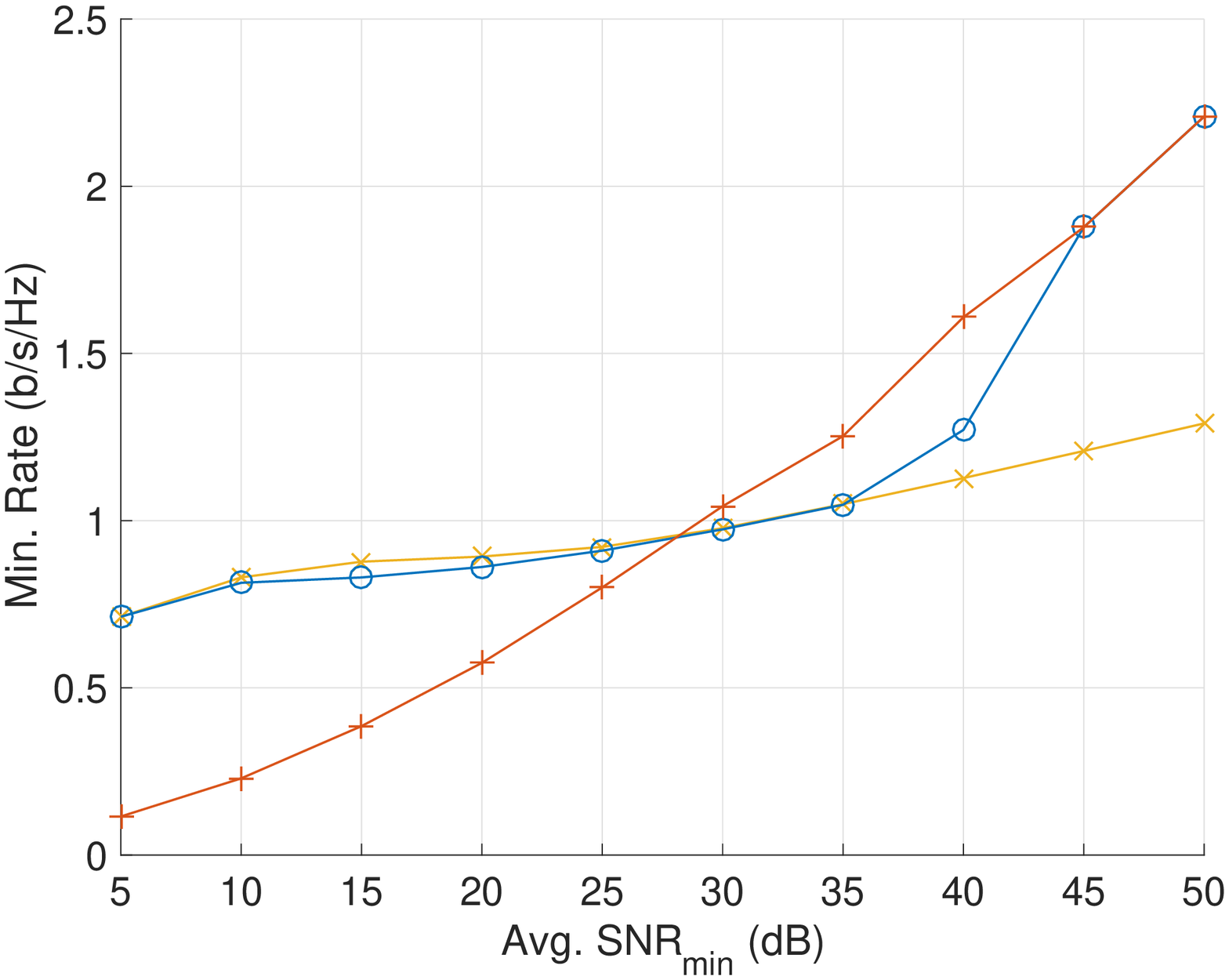}}
		\vspace{-0.15cm}
		\caption{$N=16$, $K=20$.}
		\label{fig:N16K24_PerfectCSIT_Exhaustive}
	\end{subfigure}
	\begin{subfigure}{.24\textwidth}
		\centerline{\includegraphics[width=1.9in,height=1.9in,keepaspectratio]{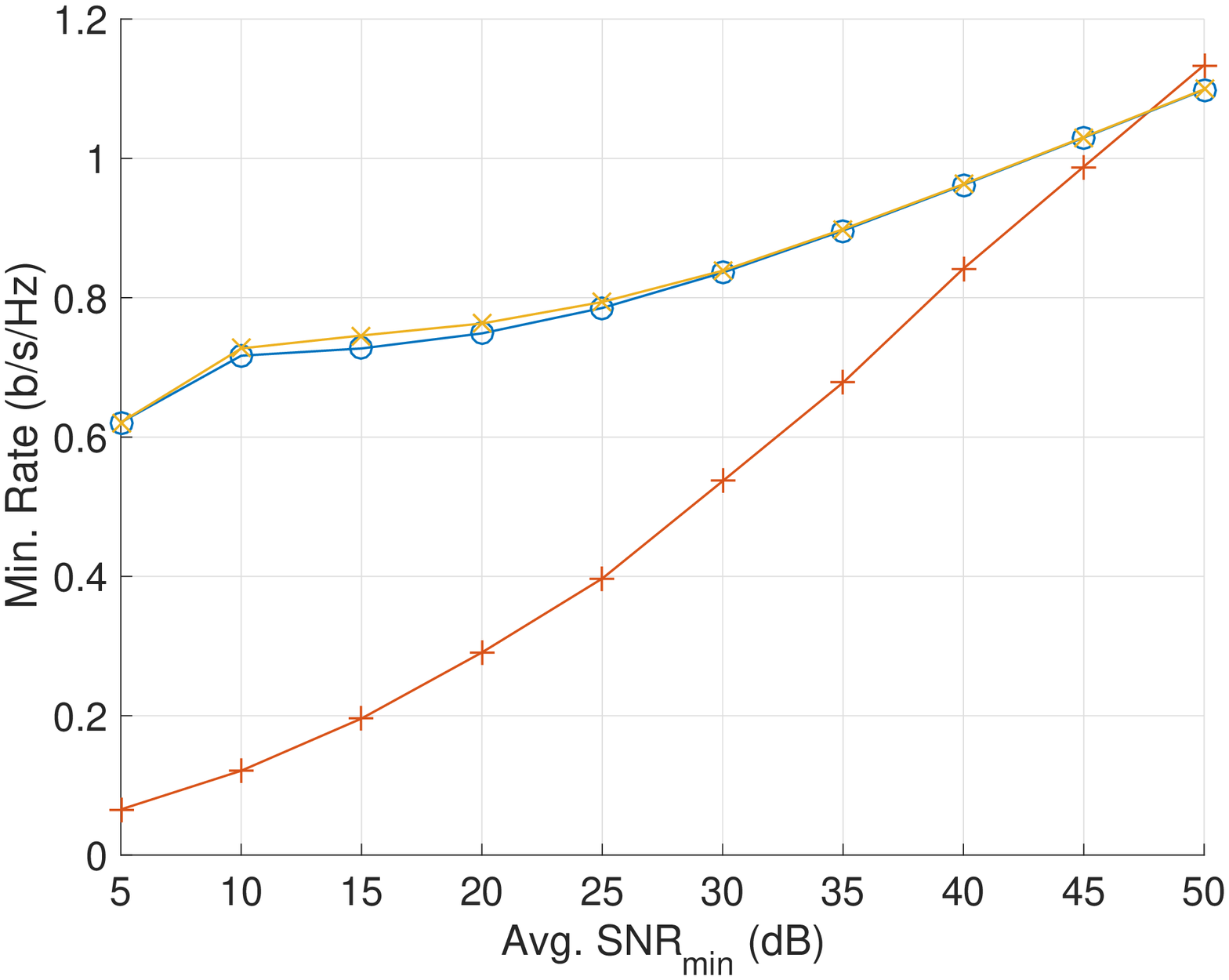}}
		\vspace{-0.15cm}
		\caption{$N=16$, $K=24$.}
		\label{fig:N32K64_PerfectCSIT_Exhaustive}
	\end{subfigure}
	\vspace{-0.6cm}
	\caption{Performance comparison with exhaustive search.}
	\vspace{-0.2cm}
\end{figure}

Next, consider the case where $R^{\mathrm{ZF}}_{\tilde{k}}(t)$ is small enough, such that, $\log_{2}\hspace{-0.01cm}(\hspace{-0.04cm}1\hspace{-0.1cm}+\hspace{-0.1cm}\sigma^{\mathrm{ZF}}_{\tilde{k}}t^{*}\hspace{-0.0cm})\hspace{-0.1cm}\approx\hspace{-0.1cm}\sigma^{\mathrm{ZF}}_{\tilde{k}}t^{*}$. We use $\log_{2}\hspace{-0.01cm}(\hspace{-0.01cm}1\hspace{-0.1cm}-\hspace{-0.1cm}\rho^{\mathrm{ZF}}\hspace{-0.1cm}+\hspace{-0.1cm}\rho^{\mathrm{ZF}}/t^{*})\approx\log_{2}(\rho^{\mathrm{ZF}}/t^{*})$ and set $\beta^{*}=0$ to rewrite \eqref{eqn:perfectcsit_highsnr} as   \vspace{-0.2cm}
\begin{align}
	\rho^{\mathrm{ZF}}&=t^{*}e^{\log(2)(K-N)\sigma^{\mathrm{ZF}}_{\tilde{k}}t^{*}}. 
	\label{eqn:perfectcsit_highsnr_assump2}
\end{align}
We multiply both sides of \eqref{eqn:perfectcsit_highsnr_assump2} by $\delta^{(2)}\triangleq\log(2)(K-N)\sigma^{\mathrm{ZF}}_{\tilde{k}}$ to obtain $\delta^{(2)}\rho^{\mathrm{ZF}}=\delta^{(2)}t^{*}e^{\delta^{(2)}t^{*}}$, which can be re-arranged as $
\delta^{(2)}t^{*}=W_{k}(\delta^{(2)}\rho^{\mathrm{ZF}})$,
where $W_{k}(x)$ is the $k$-th branch of Lambert-W function. As $\delta^{(2)}\rho^{\mathrm{ZF}}$ and $\delta^{(2)}t^{*}$ are real and $\delta^{(2)}\rho^{\mathrm{ZF}}>0$, it suffices to check only $W_{0}(x)$, so that \vspace{-0.2cm}
\begin{align}
	\delta^{(2)}t^{*}=W_{0}(\delta^{(2)}\rho^{\mathrm{ZF}}).
	\label{eqn:perfectcsit_highsnr_assump2_lambert2}
\end{align}
We use the approximation \mbox{$W_{0}(x)\approx\log(x)-\log(\log(x))$} for  large $x$ \cite{corless_1996} in \eqref{eqn:perfectcsit_highsnr_assump2_lambert2} to obtain  
\begin{align}
t^{*}=\frac{\log\left(\delta^{(2)}\rho^{\mathrm{ZF}}\right)-\log\left(\log\left(\delta^{(2)}\rho^{\mathrm{ZF}}\right)\right)}{\delta^{(2)}}\triangleq t^{(2)}.
\end{align}
Accordingly, we define the optimal rates and parameters for this scenario as in \eqref{eqn:2_perfect}.

\begin{table*}
\begin{align} 
		&t^{(2)}\hspace{-0.1cm}\triangleq\hspace{-0.1cm}
		\begin{cases}
    		\hspace{-0.1cm}\frac{\log\left(\delta^{(2)}\rho^{\mathrm{ZF}}\right)-\log\left(\log\left(\delta^{(2)}\rho^{\mathrm{ZF}}\right)\right)}{\delta^{(2)}}, & \hspace{-0.2cm}\text{if $\delta^{(2)}\rho^{\mathrm{ZF}} \geq e$} \\
    		\hspace{1.85cm}1\hspace{1.85cm}, &\hspace{-0.2cm} \text{otherwise}.
    	\end{cases}\hspace{-0.1cm}, \ \beta^{(2)}=0, \quad r^{(2)}_{mm}\hspace{-0.1cm}\triangleq\hspace{-0.1cm}\min\hspace{-0.1cm}\left\lbrace\hspace{-0.1cm}\frac{\log_{2}\hspace{-0.1cm}\left(1\hspace{-0.1cm}-\hspace{-0.1cm}\rho^{\mathrm{ZF}}\hspace{-0.1cm}+\hspace{-0.1cm}\frac{\rho^{\mathrm{ZF}}}{t^{(2)}}\right)}{K-N}, \log_{2}\hspace{-0.1cm}\left(\hspace{-0.1cm}1\hspace{-0.1cm}+\hspace{-0.1cm}\sigma^{\mathrm{ZF}}_{\tilde{k}}t^{(2)}\hspace{-0.05cm}\right)\hspace{-0.1cm}\right\rbrace\hspace{-0.1cm}.	
	\label{eqn:2_perfect}
\end{align}
\vspace{-0.15cm}
\hrule
\vspace{-0.3cm}
\end{table*}

\subsection{Solution Using MRT Precoding}
\label{sec:solution_perfectcsit_MRT}
\vspace{-0.1cm}
Different from the design in the previous section, each user is served by at least one private stream, so that, allocating power to the common stream is not mandatory to achieve a non-zero rate for each user, which yields $t \leq 1$.
Again, we solve the problem by using the lower bounds derived in Section~\ref{sec:boundRate}. 
Using \eqref{eqn:prop4}, we redefine $\hat{k}=\argmin_{k \in \mathcal{K}}v_{k}$, such that, $R^{\mathrm{MRT}}_{\hat{k}}(t)=\log_{2}(1+\alpha^{\mathrm{MRT}}t)-\log_{2}(1+\lambda^{\mathrm{MRT}}t)$, where $\alpha^{\mathrm{MRT}}=v_{\hat{k}}\frac{P}{K}e^{\psi(N+K-1)}$, $\lambda^{\mathrm{MRT}}=v_{\hat{k}}\frac{P(K-1)}{K}$ for given $v_{k}$, $\forall k \in \mathcal{K}$. 
Combining with \eqref{eqn:lemma3}, we can rewrite \eqref{eqn:MRTproblem} as
\begin{align}
	\max_{t\in (0,1]}\hspace{-0.09cm}\frac{1}{K}\log_{2}\hspace{-0.09cm}\left(\hspace{-0.1cm}1\hspace{-0.05cm}-\hspace{-0.05cm}\rho^{\mathrm{MRT}}\hspace{-0.1cm}+\hspace{-0.1cm}\frac{\rho^{\mathrm{MRT}}}{t}\hspace{-0.05cm}\right)\hspace{-0.1cm}+\hspace{-0.1cm}\log_{2}\hspace{-0.05cm}\left(\frac{1+\alpha^{\mathrm{MRT}}t}{1+\lambda^{\mathrm{MRT}}t}\hspace{-0.05cm}\right)\hspace{-0.05cm}.
	\label{eqn:MRTproblem_simple_2}
\end{align}
Assuming $\log_{2}(1-\rho^{\mathrm{MRT}}+\rho^{\mathrm{MRT}}/t^{*})\approx\log_{2}(\rho^{\mathrm{MRT}}/t^{*})$ for simplicity.
Taking the derivative of \eqref{eqn:MRTproblem_simple_2} with this approximation and equating the result to zero, we obtain \vspace{-0.1cm}
\begin{align}
	&\underbrace{-\alpha^{\mathrm{MRT}}\hspace{-0.1cm}\lambda^{\mathrm{MRT}}}_{a}(t^{*})^{2}\hspace{-0.1cm}+\hspace{-0.1cm}\underbrace{\left[\hspace{-0.05cm}K(\alpha^{\mathrm{MRT}}\hspace{-0.15cm}-\hspace{-0.1cm}\lambda^{\mathrm{MRT}})\hspace{-0.1cm}-\hspace{-0.1cm}(\alpha^{\mathrm{MRT}}\hspace{-0.15cm}+\hspace{-0.1cm}\lambda^{\mathrm{MRT}})\right]}_{b}\hspace{-0.1cm}t^{*}\hspace{-0.1cm}=\hspace{-0.1cm}1. \nonumber
\end{align}
Then, one can write the roots $s_{1/2}=\frac{-b\pm\sqrt{b^{2}+4a}}{2a}$. In order to have a real-valued $t^{*}\geq 0$, the conditions $\sqrt{b^{2}+4a}\geq 0$ and $b\geq 0$ should be satisfied. Accordingly, the optimal rates and parameters for this scenario are given in \eqref{eqn:45_perfect}. 
\begin{table*}
\begin{align} 
		&t^{(3/4)}\triangleq
		\begin{cases}
    		\min\left\lbrace1,\frac{-b\pm\sqrt{b^{2}+4a}}{2a}\right\rbrace, & \hspace{-0.2cm}\text{if $\sqrt{b^{2}+4a} \geq 0$, $b \geq 0$,} \\
    		\hspace{1.4cm}1\hspace{1.4cm}, &\hspace{-0.2cm} \text{otherwise},
    	\end{cases}, \quad r^{(3/4)}_{mm}\hspace{-0.1cm}\triangleq\hspace{-0.1cm}\frac{1}{K}\hspace{-0.05cm}\log_{2}\hspace{-0.1cm}\left(\hspace{-0.1cm}1\hspace{-0.1cm}-\hspace{-0.1cm}\rho^{\mathrm{MRT}}\hspace{-0.15cm}+\hspace{-0.15cm}\frac{\rho^{\mathrm{MRT}}}{t^{(3/4)}}\hspace{-0.1cm}\right)\hspace{-0.1cm}+\hspace{-0.1cm}\log_{2}\hspace{-0.1cm}\left(\hspace{-0.1cm}\frac{1\hspace{-0.1cm}+\hspace{-0.1cm}\alpha^{\mathrm{MRT}}t^{(3/4)}}{1\hspace{-0.1cm}+\hspace{-0.1cm}\lambda^{\mathrm{MRT}}t^{(3/4)}}\hspace{-0.1cm}\right)\hspace{-0.1cm}.\hspace{-0.15cm}	
	\label{eqn:45_perfect}
\end{align}
\vspace{-0.2cm}
\hrule
\vspace{-0.3cm}
\end{table*}
\begin{figure}[t!]
	\begin{subfigure}{.24\textwidth}
		\centerline{\includegraphics[width=1.9in,height=1.9in,keepaspectratio]{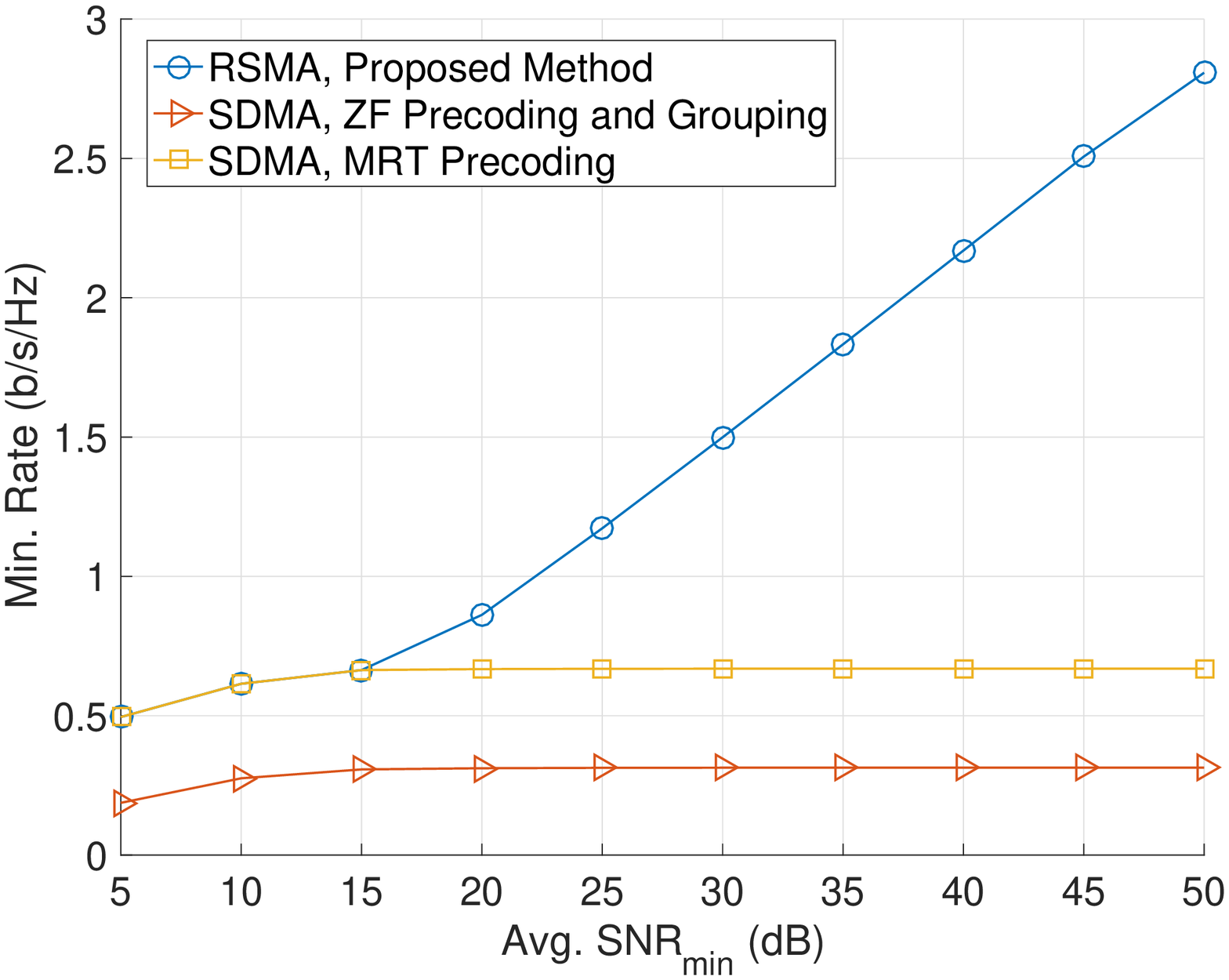}}
		\vspace{-0.15cm}
		\caption{$N=4$, $K=8$.}
		\label{fig:N4K8_PerfectCSIT_Benchmarks}
	\end{subfigure}
	\begin{subfigure}{.24\textwidth}
		\centerline{\includegraphics[width=1.9in,height=1.9in,keepaspectratio]{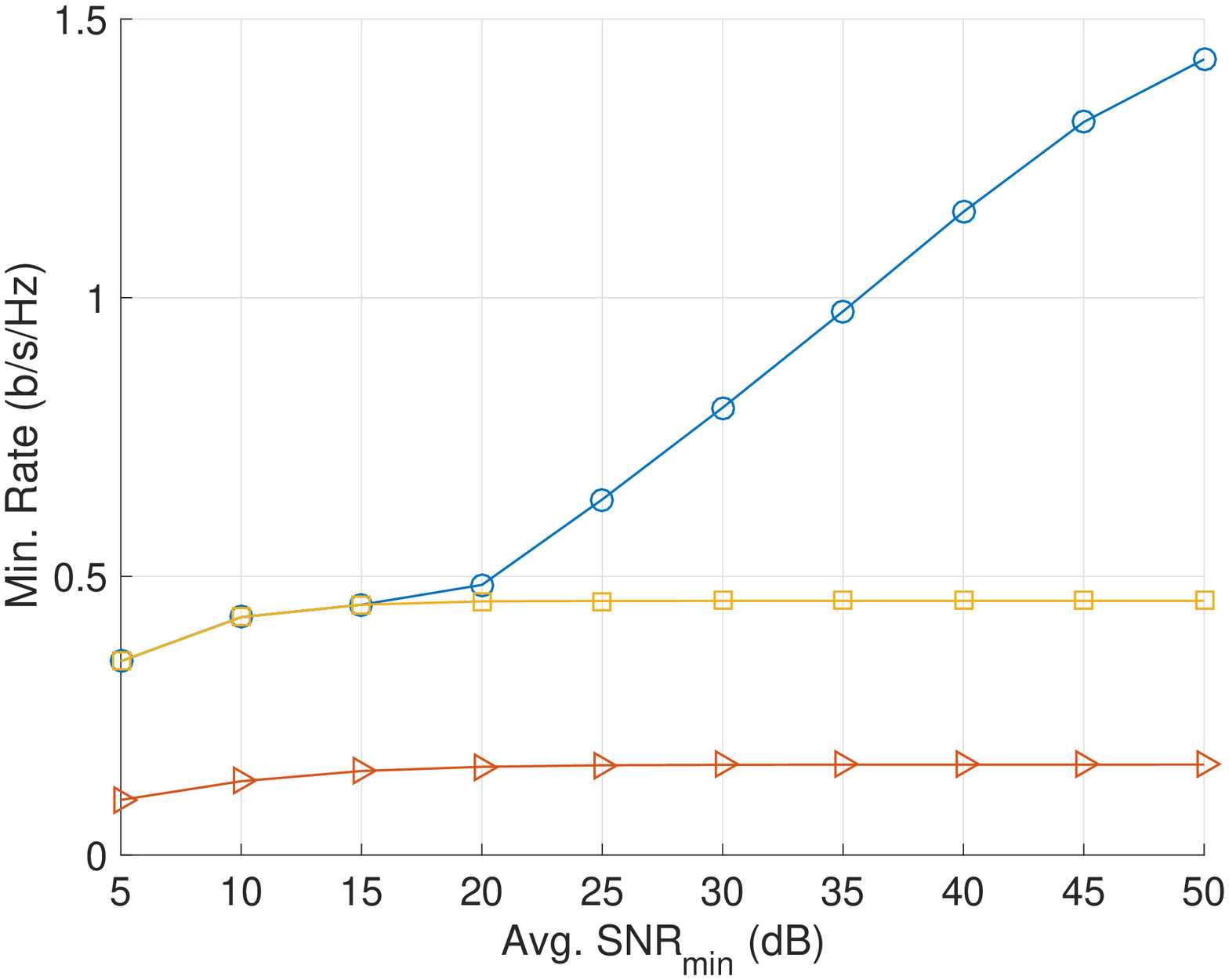}}
		\vspace{-0.15cm}
		\caption{$N=4$, $K=12$.}
		\label{fig:N16K24_PerfectCSIT_Benchmarks}
	\end{subfigure}
	\newline
	\begin{subfigure}{.24\textwidth}
		\centerline{\includegraphics[width=1.9in,height=1.9in,keepaspectratio]{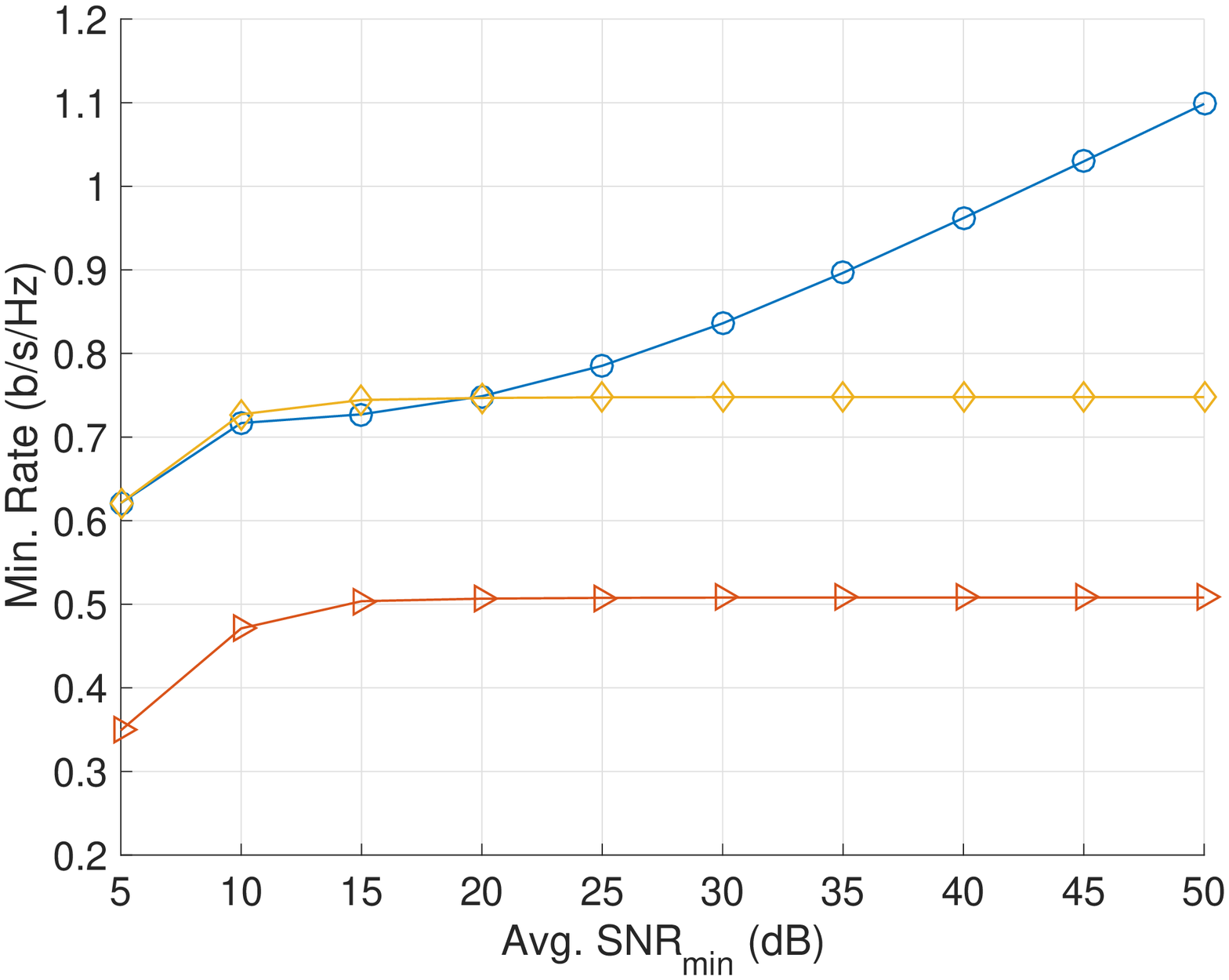}}
		\vspace{-0.15cm}
		\caption{$N=16$, $K=32$.}
		\label{fig:N16K24_PerfectCSIT_Benchmarks}
	\end{subfigure}
	\begin{subfigure}{.24\textwidth}
		\centerline{\includegraphics[width=1.9in,height=1.9in,keepaspectratio]{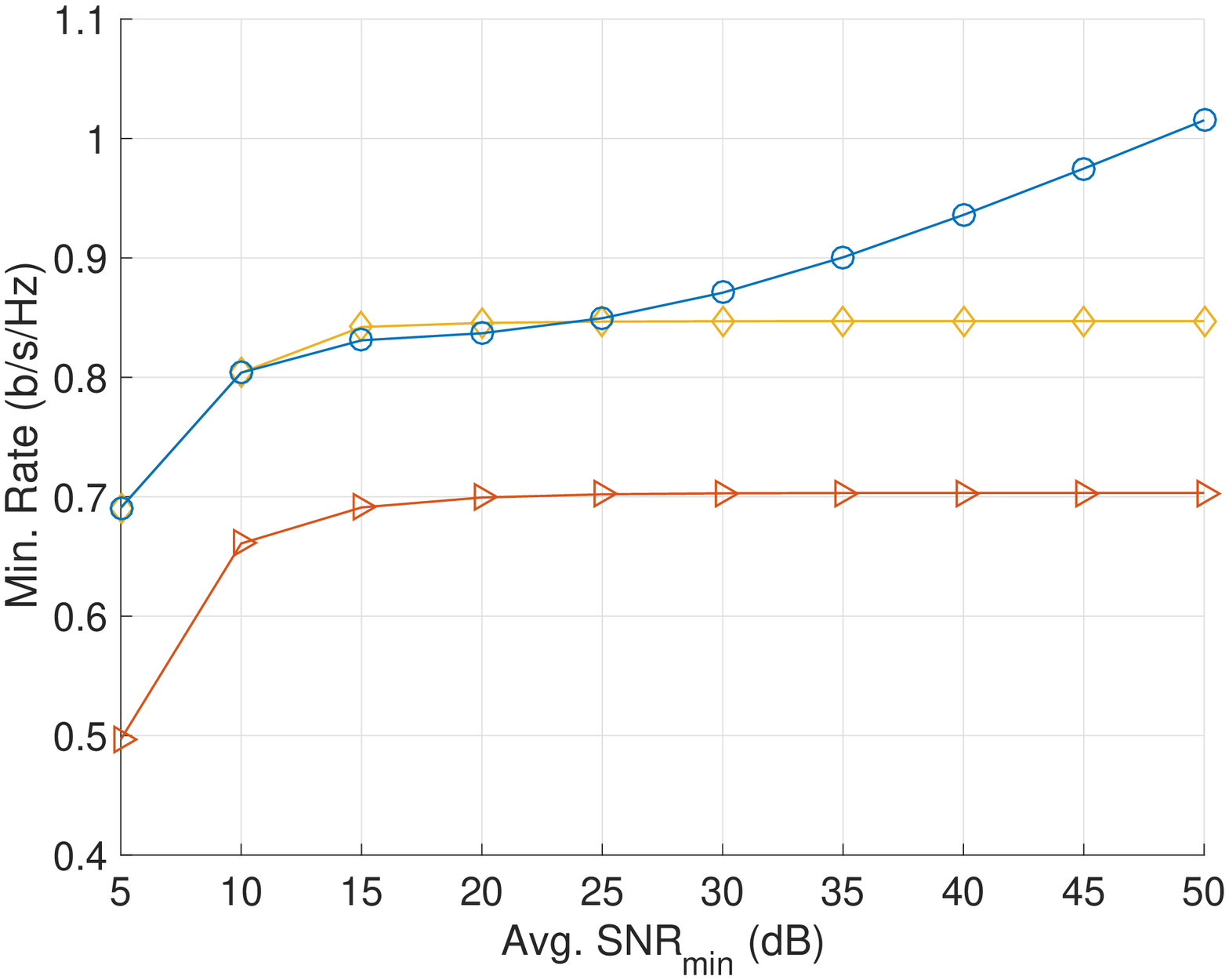}}
		\vspace{-0.15cm}
		\caption{$N=32$, $K=40$.}
		\label{fig:N32K40_PerfectCSIT_Benchmarks}
	\end{subfigure}
	\vspace{-0.6cm}
	\caption{Performance comparison with benchmark schemes.}
	\vspace{-0.3cm}
\end{figure}
\subsection{Proposed Precoder and Resource Allocation}
\label{sec:perfectcsit_proposed}
Finally, we describe the proposed power and precoder allocation method. The optimal power allocation $t_{opt}$, rate allocation $\beta_{opt}$, and precoder allocation for the private stream of user-$k$, $\mathbf{p}_{opt,k}$, are performed as \vspace{-0.2cm}
\begin{align}
	&\hat{n}=\argmax_{n \in \{1,\ldots,4\}}r^{(n)}_{mm}, \ \mathbf{p}^{*}_{k,opt}\hspace{-0.1cm}=\hspace{-0.1cm} 
	\begin{cases}
	 	\mathbf{p}^{\mathrm{ZF}}_{k}\hspace{0.25cm}, &\hspace{-0.3cm} \text{if $\hat{n} \leq 2$ and $k \in \mathcal{G}_{1}$,} \\
	 	\hspace{0.45cm}\mathbf{0}\hspace{0.45cm}, & \hspace{-0.3cm}\text{if $\hat{n} \leq 2$ and $k \in \mathcal{G}_{2}$,} \\
	 	\mathbf{p}^{\mathrm{MRT}}_{k}, & \hspace{-0.3cm}\text{if $\hat{n} \geq 3$ and $k \in \mathcal{K}$,} \\
	\end{cases}\nonumber \\
	&t_{opt}=t^{(\hat{n})},\quad \beta_{opt}=0.
	\label{eqn:proposed_perfect}
\end{align}
The common precoder, $\mathbf{p}_{c,opt}$, can be calculated as the left-most eigenvector of the channel matrix $\mathbf{H}=\left[\mathbf{h}_{1}, \mathbf{h}_{2}, \ldots, \mathbf{h}_{K}\right]$.

\section{Numerical Results}
\label{sec:numerical}
We demonstrate the performance of the proposed scheme with numerical results. We perform Monte Carlo simulations over $100$ different channel realizations to investigate the max-min ergodic rate performance.    
First, we compare the performance of the proposed precoder and resource allocation method in \eqref{eqn:proposed_perfect} with the performance of resource allocation by exhaustive search. The exhaustive search calculates the rate values $r^{(n)}_{mm}$ for all possible values of $t$ and $\beta$ with a certain resolution and choose the values that maximize the minimum rate. The search is performed over $t \in [10^{-6}, 1]$ and $\beta \in [0,1/K]$ with different resolutions in different intervals to reduce the complexity, so that $130$ values are considered for $t$ and $\lceil\frac{1}{0.001K}\rceil$ values are considered for $\beta$.

Fig.~\ref{fig:N4K6_PerfectCSIT_Exhaustive}-\ref{fig:N32K64_PerfectCSIT_Exhaustive} show the minimum rate performance comparison between the proposed method and exhaustive search with perfect and imperfect CSIT for $N=4$, $K=6$ and $N=4$, $K=8$. The x-axis of the figures represent the average SNR of the user with minimum large scale fading coefficient, {\sl i.e.}, $\min_{k \in \mathcal{K}}v_{k}P$. The values of $v_{k}$ are independently chosen by uniform distribution between $0.1$ and $1$, {\sl i.e.}, $v_{k}\sim \mathrm{U}[0.1,1]$, with at least one user satisfying $v_{k}=0.1$ and one user satisfying $v_{k}=1$. The user channels $\mathbf{h}_{k}$ are independent r.v.s with Rayleigh distribution and unit variance. For simplicity, we determine $G_{1}$ for ZF precoding according to the user indexes, {\sl i.e.}, $\mathcal{G}_{1}={1, 2, \ldots, N}$. As one can observe from the figures, the proposed method performs very close to the optimal performance and can switch between ZF and MRT modes effectively. This verifies that the proposed method achieves close to optimal performance with low complexity. 

Next, we compare the performance of RSMA with the proposed algorithm with other benchmark schemes. For comparison, we consider one-shot transmission schemes with no time-sharing. Such schemes are suitable for systems with tight scheduling and latency constraints \cite{yin_2021}. We consider the following 2 benchmark schemes:
\begin{enumerate}
	\item \textit{SDMA with ZF precoding and grouping:} Users are divided into $\lceil\frac{K}{N}\rceil$ groups according to their user indexes and each user calculates its ZF precoder to null interference in the corresponding group. 
	\item \textit{SDMA with MRT precoding:} Users are served by MRT.  
\end{enumerate}

Fig.~\ref{fig:N4K8_PerfectCSIT_Benchmarks}-\ref{fig:N64K80_PerfectCSIT_Benchmarks} show the performance comparison between the proposed method and the benchmark schemes for various $N$, $K$. From the figures, one can observe the rate saturation for SDMA with respect to SNR due to multi-user interference.
On the other hand, RSMA achieves a non-saturating rate for every considered scenario. Therefore, we can conclude that RSMA achieves a significant performance gain over SDMA-based schemes in overloaded MIMO systems.  

\section{Conclusion and Future Work}
\label{sec:conclusion}
We study RSMA in overloaded downlink MIMO networks with perfect CSIT. We formulate a max-min fairness problem to find the optimal precoder and resource allocation. We provide a low-complexity solution for the formulated problem by using ZF and MRT precoders and deriving closed-form solutions for rate and power allocation between the common and private streams. We demonstrate that the proposed low-complexity design achieves a near-optimal performance. Numerical results also show that the proposed design achieves a non-saturating rate performance as opposed to SDMA. Future work includes improving the performance of the proposed method at medium/low SNR and analysis and design under imperfect CSIT.
\appendices
\begin{table*}[t!]
%\vspace{-0.2cm}
	\begin{align}
		\frac{-N}{K-N}\frac{1}{\log(2)}\log\left(1-\rho+(\rho\sigma^{\mathrm{ZF}}_{\tilde{k}})^{\frac{K-N}{1-K\beta+K-N}}\right) + \frac{1-N\beta}{K-N}\frac{1}{\log(2)}\frac{(\rho\sigma^{\mathrm{ZF}}_{\tilde{k}})^{\frac{K-N}{1-K\beta+K-N}}\log(\rho\sigma^{\mathrm{ZF}}_{\tilde{k}})\frac{N(K-N)}{(1-K\beta+K-N)^{2}}}{1-\rho+\left(\rho\sigma^{\mathrm{ZF}}_{\tilde{k}}\right)^{\frac{K-N}{1-K\beta+K-N}}}.
		\label{eqn:lemmax_derivative} 
	\end{align}
	 \vspace{-0.15cm}
	\hrule
\end{table*}
\begin{table*}
\vspace{-0.2cm}
	\begin{align}
		\left(1-\rho+\left(\rho\sigma^{\mathrm{ZF}}_{\tilde{k}}\right)^{\frac{K-N}{1-K\beta+K-N}}\right)&\log\left(1-\rho+\left(\rho\sigma^{\mathrm{ZF}}_{\tilde{k}}\right)^{\frac{K-N}{1-K\beta+K-N}}\right)> \left(\rho\sigma^{\mathrm{ZF}}_{\tilde{k}}\right)^{\frac{K-N}{1-K\beta+K-N}}\log\left(\rho\sigma^{\mathrm{ZF}}_{\tilde{k}}\right)\frac{(1-N\beta)(K-N)}{(1-K\beta+K-N)^{2}}.
		\label{eqn:lemmax_derivative_rearranged} 	
	\end{align}
	\vspace{-0.15cm}
	\hrule
	\vspace{-0.6cm}
\end{table*}
\section{Proof of Proposition 1}
\label{appendix:prop1}
We define the random variables (r.v.s) $X^{\mathrm{ZF}}_{k}=v_{k}|\mathbf{h}_{k}^{H}\mathbf{p}^{\mathrm{ZF}}_{k}|^{2}$, $Y^{\mathrm{ZF}}=\min_{k\in\mathcal{K}}Y^{\mathrm{ZF}}_{k}$, and\vspace{-0.1cm}
\begin{align} 
Y^{\mathrm{ZF}}_{k}=
\begin{cases}
	\frac{v_{k}|\mathbf{h}_{k}^{H}\mathbf{p}_{c}|^{2}}{1+\frac{Pt}{N}v_{k}|\widehat{\mathbf{h}}_{k}^{H}\mathbf{p}^{\mathrm{ZF}}_{k}|^{2}}\hspace{0.95cm}, & \hspace{-0.2cm}\text{if $k \in \mathcal{G}_{1}$,} \\ 
	\frac{v_{k}|\mathbf{h}_{k}^{H}\mathbf{p}_{c}|^{2}}{1+\frac{Pt}{N}v_{k}\sum_{j \in \mathcal{G}_{1}}|\mathbf{h}_{k}^{H}\mathbf{p}^{\mathrm{ZF}}_{j}|^{2}}, & \hspace{-0.2cm}\text{if $k \in \mathcal{G}_{2}$,}
	\end{cases} 
	\label{eqn:rvs} 
\end{align}
such that, 
\begin{align}
	R^{\mathrm{ZF}}_{c}(t)&=
	\mathbb{E}\left\lbrace \log_{2}\hspace{-0.1cm}\left(1\hspace{-0.1cm}+\hspace{-0.1cm}P(1-t)Y^{\mathrm{ZF}} \right) \hspace{-0.1cm}\right\rbrace, \nonumber \\
	R^{\mathrm{ZF}}_{k}(t)&=\mathbb{E}\left\lbrace \log_{2}\hspace{-0.05cm}\left( 1\hspace{-0.1cm}+\hspace{-0.1cm} \frac{Pt}{N}X^{\mathrm{ZF}}_{k} \right) \right\rbrace.
\end{align}
The proof follows the steps of \cite[Proposition 2]{dizdar_2021} by finding the CDF of $Y^{\mathrm{ZF}}_{k}$ for the users in $\mathcal{G}_{1}$ and $\mathcal{G}_{2}$ separately, using them to find the CDF of $Y^{\mathrm{ZF}}$ and applying \cite[Lemma 4]{dizdar_2021}. We skip the details here for the sake of brevity.% and refer the interested reader to \cite{XXX}.     

%\vspace{-0.2cm}
%We start by deriving the CDF of $Y^{\mathrm{ZF}}$. Benefiting from \cite[Lemma 2]{dizdar_2021} and \eqref{eqn:rvs}, $Y^{\mathrm{ZF}}$ can be approximated by $\widetilde{Y}^{\mathrm{ZF}}$ with CDF $F_{\widetilde{Y}^{\mathrm{ZF}}}(y)=
%	1-\frac{e^{-y\sum_{k=1}^{N}\frac{1}{v_{k}}}}{\left(y\frac{Pt}{N}+1 \right)^{N}}\frac{e^{-y\sum_{k=N+1}^{K}\frac{1}{v_{k}}}}{\left(y\frac{Pt}{N}+1 \right)^{N(K-N)}}$. 
%Let us define $\widetilde{R}^{\mathrm{ZF}}_{c}(t)=\log_{2}\left( 1+P(1-t) \widetilde{Y}^{\mathrm{ZF}}\right)$. We can write:  \vspace{-0.2cm}
%\begin{align}
%	R^{\mathrm{ZF}}_{c}(t)\approx\widetilde{R}^{\mathrm{ZF}}_{c}(t) \geq \log_{2}\hspace{-0.1cm}\left(\hspace{-0.1cm}1+P(1-t)e^{\mathbb{E}\left\lbrace\log\left(\widetilde{Y}^{\mathrm{ZF}}\right)\right\rbrace}\hspace{-0.1cm}\right).
%	\label{eqn:prop1_proof1}
%\end{align}
%The inequality in \eqref{eqn:prop1_proof1} follows from Jensen's inequality and the fact that $\log_{2}(1+ae^{x})$ is a convex function of $x$ for any $a>0$. 
%Next, we benefit from \cite[Lemma 3]{dizdar_2021} to obtain $\mathbb{E}\left\lbrace\log\left(\tilde{Y}^{\mathrm{ZF}}\right)\right\rbrace=-\gamma-\log\left(\sum_{k=1}^{K}\frac{1}{v_{k}}\right) -e^{\frac{N}{Pt}\sum_{k=1}^{K}\frac{1}{v_{k}}}\hspace{-0.2cm}\sum_{m=1}^{N(K-N+1)}\hspace{-0.6cm}E_{m}\left(\frac{N}{Pt}\sum_{k=1}^{K}\frac{1}{v_{k}}\right)\triangleq\eta^{\mathrm{ZF}}$ and substitute into \eqref{eqn:prop1_proof1} to complete the proof.
%\vspace{-0.2cm}
\section{Proof of Proposition 5}
\label{appendix:propx}
We prove Proposition 5 by contradiction. Let us assume that the maximum for problem \eqref{eqn:ZFproblem_2} is attained by $\beta^{*}$ and $t^{*}$ satisfying $\frac{1-N\beta^{*}}{K-N}R^{\mathrm{ZF}}_{c}(t^{*})<\beta^{*}R^{\mathrm{ZF}}_{c}(t^{*})+R^{\mathrm{ZF}}_{\tilde{k}}(t^{*})$ for $0<\beta^{*}<1/N$ and $0\leq t^{*}<1$. Then, one can find $\beta^{\prime}=\beta^{*}-\epsilon$ for arbitrarily small $\epsilon>0$, such that, $\frac{1-N\beta^{\prime}}{K-N}R^{\mathrm{ZF}}_{c}(t^{*})>\frac{1-N\beta^{*}}{K-N}R^{\mathrm{ZF}}_{c}(t^{*})$ and $\frac{1-N\beta^{\prime}}{K-N}R^{\mathrm{ZF}}_{c}(t^{*})<\beta^{\prime}R^{\mathrm{ZF}}_{c}(t^{*})+R^{\mathrm{ZF}}_{\tilde{k}}(t^{*})$, which contradicts the initial assumption.
Next, we assume that maximum is attained by $\beta^{*}=0$ and $t^{*}$ satisfying $\frac{1}{K-N}R^{\mathrm{ZF}}_{c}(t^{*})<R^{\mathrm{ZF}}_{\tilde{k}}(t^{*})$ for $0<t^{*}<1$. We note that $t$ is strictly positive in this case to obtain a non-zero maximum value under this assumption. By observing the first derivatives, it can be shown that $R^{\mathrm{ZF}}_{c}(t)$ is monotonically decreasing and $R^{\mathrm{ZF}}_{k^{\prime}}(t)$ is monotonically increasing with $t$. Consequently, we can find $t^{\prime}=t^{*}-\epsilon$ for arbitrarily small $\epsilon>0$, such that, $\frac{1}{K-N}R^{\mathrm{ZF}}_{c}(t^{\prime})>\frac{1}{K-N}R^{\mathrm{ZF}}_{c}(t^{*})$ and $\frac{1}{K-N}R^{\mathrm{ZF}}_{c}(t^{\prime})<R^{\mathrm{ZF}}_{\tilde{k}}(t^{\prime})$, which contradicts the initial assumption. 

In order to complete the proof, one can repeat the procedure above in a similar fashion for the scenarios where maximum is attained by $\beta^{*}$ and $t^{*}$ satisfying $\frac{1-N\beta^{*}}{K-N}R^{\mathrm{ZF}}_{c}(t^{*})>\beta^{*}R^{\mathrm{ZF}}_{c}(t^{*})+R^{\mathrm{ZF}}_{\tilde{k}}(t^{*})$, which is omitted here for brevity.   
The solutions $\beta^{*}$ and $t^{*}$ are unique as they are optimum solutions for a max-min problem \cite[Theorem 1.2.2]{boudec_2021}. 
\vspace{-0.1cm}
\section{Proof of Lemma 1}
\label{appendix:lemmax} 
We prove Lemma 1 by showing that the derivative of $\frac{1-N\beta}{K-N}R^{\mathrm{ZF}}_{c}(t^{(1)}(\beta))$ with respect to $\beta$, which is given in \eqref{eqn:lemmax_derivative}, is strictly negative. 
The first term in \eqref{eqn:lemmax_derivative} is strictly negative for $(\rho\sigma^{\mathrm{ZF}}_{\tilde{k}})^{\frac{K-N}{1-K\beta+K-N}}>\rho$, which is satisfied for $\sigma^{\mathrm{ZF}}_{\tilde{k}}>1$.
The second term is non-positive for $\sigma^{\mathrm{ZF}}_{\tilde{k}}\rho\leq1$, which results in the overall derivative to be negative. For the case $\sigma^{\mathrm{ZF}}_{\tilde{k}}\rho>1$, we need to show that $N\log\left(1-\rho+\left(\rho\sigma^{\mathrm{ZF}}_{\tilde{k}}\right)^{\frac{K-N}{1-K\beta+K-N}}\right) > \frac{(1-N\beta)\left(\rho\sigma^{\mathrm{ZF}}_{\tilde{k}}\right)^{\frac{K-N}{1-K\beta+K-N}}\log\left(\rho\sigma^{\mathrm{ZF}}_{\tilde{k}}\right)\frac{(K-N)}{(1-K\beta+K-N)^{2}}}{1-\rho+\left(\rho\sigma^{\mathrm{ZF}}_{\tilde{k}}\right)^{\frac{K-N}{1-K\beta+K-N}}}$. For this purpose, we rearrange the inequality as in \eqref{eqn:lemmax_derivative_rearranged}.
Since $1-\rho \geq 0$, the first term in \eqref{eqn:lemmax_derivative_rearranged} is lower bounded by %\eqref{eqn:lemmax_lb}.
%\begin{table*}
\vspace{-0.3cm}
	\begin{align}
	%&\left(1-\rho+\left(\rho\sigma^{\mathrm{ZF}}_{\tilde{k}}\right)^{\frac{K-N}{1-K\beta+K-N}}\right)\log\left(\left(\rho\sigma^{\mathrm{ZF}}_{\tilde{k}}\right)^{\frac{K-N}{1-K\beta+K-N}}\right)>\left(\rho\sigma^{\mathrm{ZF}}_{\tilde{k}}\right)^{\frac{K-N}{1-K\beta+K-N}}\frac{(K-N)\log\left(\rho\sigma^{\mathrm{ZF}}_{\tilde{k}}\right)}{1-K\beta+K-N}.
	\left(\rho\sigma^{\mathrm{ZF}}_{\tilde{k}}\right)^{\frac{K-N}{1-K\beta+K-N}}\frac{(K-N)\log\left(\rho\sigma^{\mathrm{ZF}}_{\tilde{k}}\right)}{1-K\beta+K-N}.
	\label{eqn:lemmax_lb}
	\end{align}
	%\vspace{-0.15cm}
%	\hrule
%\end{table*}
Applying the lower bound \eqref{eqn:lemmax_lb} for the first term in \eqref{eqn:lemmax_derivative_rearranged} and cancelling out the identical terms on both sides, we can obtain $\frac{(1-N\beta)}{(1-K\beta+K-N)} < 1$, 
which holds since $K-N>1$ and $\beta \leq \frac{1}{K}$. Therefore, we can conclude that $\frac{\partial}{\partial\beta}\frac{1-N\beta}{K-N}R^{\mathrm{ZF}}_{c,p}(t^{(1)}_{p}(\beta)) <0$ and $\frac{1-N\beta}{K-N}R^{\mathrm{ZF}}_{c,p}(t^{(1)}_{p}(\beta))$ is monotonically decreasing with $\beta$. 
%\section*{References}

\end{document}